\documentclass[12pt,a4papper]{article}

\usepackage{graphics}
\usepackage{graphicx}
\begin{document}

\pagestyle{myheadings}
\markright{{\bf A.F. Ra\~nada, Cosmological variation of {\mathversion{bold}$\alpha$}}}
\title{On the cosmological variation of the fine structure constant}

\author{Antonio F. Ra\~nada\\Departamento de F\'{\i}sica Aplicada III,
Universidad Complutense, \\28040 Madrid, Spain\thanks{E-mail: afr@fis.ucm.es}}

\date{November 1, 2001}
\maketitle

\begin{abstract}
A phenomenological model is proposed to explain the recent observed cosmological variation of the fine structure
constant as an effect of the quantum vacuum, assuming a flat universe with cosmological constant $\Lambda$ in the cases $(\Omega _M, \Omega _\Lambda )$ equal to $(0.3, 0.7)$ and $(1,0)$. Because of the fourth Heisenberg relation, the lifetime of
the virtual pairs of the zero-point radiation must depend on the gravitational potential $\Phi$, so that the
quantum vacuum changes its density and acquires a relative permittivity different from one. Since the matter was
more concentrated in the past, the gravitational potential of all the universe was stronger and the optical
density of the vacuum higher, the electron charge being therefore more renormalized and smaller than now. The
model is based on a first order Newtonian approximation that is valid for the range of the observations, but not
for very high redshift, the prediction being that $\Delta \alpha /\alpha$ is proportional to $\{\Omega
_M[a(t)^{-1}-1]-2\Omega _\Lambda [a(t)^2-1]\}$, $a(t)$ being the scale factor. This agrees with the
observations.
\end{abstract}

PACS numbers: 06.20.Jr, 95.30Sf, 98.80.Es

Short title: Cosmological variation of $\alpha$

\newpage

{\bf Introduction}. The observations of the absorption lines of distant quasars by
Webb {\em et al} seem to indicate that the fine structure constant was smaller
in the past \cite{Web01}.
 The idea that the fundamental constants are actually changing goes back to the ``large number hypothesis" by Dirac \cite{Dir37}, what stimulated a number of proposals to formulate the variation of the constants, see for instance ref. \cite{Bar02} in which $\alpha$ is written in terms of a scalar field, or other theories in which the variation of $\alpha$ is induced by a variation of the light velocity \cite{Bar99} (see also \cite{Chu98}).

This letter proposes a way to understand
this phenomenon as an effect of the quantum vacuum. More precisely,
it is argued here that, because of the fourth Heisenberg relation, the density of the
sea of virtual particles in the quantum vacuum must change
in a gravitational field, with a corresponding variation of its permittivity
and permeability. As a consequence the quantum vacuum was optically denser in the past, when the universe was more contracted than now and the gravitational potential due to all the universe was stronger.  The observed electron charge $e$ is the result of
the renormalization of its bare charge $e_{\rm bare}(>e)$ by the quantum vacuum \cite{Mil94},
so that a denser vacuum implies a smaller observed charge. It would be desirable to study this question in the frame of a rigorous quantum field
theory, but the quantum vacuum is not understood well enough today in order to do that (for instance it is not known why its energy density seems to be so small). The only available alternative (other than waiting for a future theory) is to try a phenomenological approach, as is done here.

In the model here proposed, the quantum vacuum is treated as a transparent optical medium characterized by its permittivity and permeability, the change of $\alpha$ being a consequence of the fourth Heisenberg relation applied to the gravitational interaction of the virtual pairs in the zero point radiation with all the universe. The analysis is  nonrelativistic (although with the inclusion of the rest-mass energy $mc^2$) and only weak Newtonian gravitational potentials $\Phi$ are considered for which $|\Phi|/c^2\ll 1$.

{\bf The fourth Heisenberg relation and the vacuum density}. Traditionally, quantum physics has stated that the sea of virtual pairs that
 are  created and destroyed constantly in the quantum vacuum, {\em i. e.} the
zero-point energy, has infinite density,
as follows from the simple application of its basic principles. However, there is now evidence that this density may be finite.
On the average and phenomenologically, a virtual pair created with energy $E$ (including rest-mass energy,
kinetic energy and electromagnetic energy) will live during
a time $\tau _0=\hbar /E$, according to the fourth Heisenberg relation.  A
constant number density of pairs is established in this way at a certain
equilibrium between the number of particles created and destroyed
per unit time. This has an important consequence: virtual pairs must
live longer in the gravitational field created by a mass distribution, because they have  an extra negative
potential energy $E\Phi /c^2$ there ($\Phi$ being the Newtonian potential).
Indeed their lifetime must be
\begin{equation}
\tau _\Phi =\hbar/(E+E\Phi /c^2)= \tau _0/(1+\Phi /c^2).
\label{10}
\end{equation}
The consequence is clear if unexpected: the number density of pairs
$\cal N$ depends on the gravitational potential $\Phi$ as
\begin{equation}
{\cal N}_\Phi ={\cal N}_0/ (1+\Phi /c^2).
\label{20}
\end{equation}
In other words, the density of the quantum vacuum depends  on the
gravitational potential because it must have there an extra number
density of pairs depending on $\Phi/c^2$. If $\Phi <0$, the
quantum vacuum becomes denser, if $\Phi >0$, it becomes thinner.
These two cases correspond to the gravity created by the mass and
the cosmological constant, respectively. However, this change of
the number density can produce no gravitational effects, since the
gravitational mass of the virtual pairs in a volume depends on
$\Phi$ through the product of their mass $E/c^2$ by the number
density per unit volume and unit energy ${\cal N}(E)$. This
product does not change since the second factor acquires a factor
$(1+\Phi /c^2)^{-1}$ in a gravitational field, the first a factor
$(1+\Phi /c^2)$, as we have seen. The effect that we are
considering is optical and dielectric, not gravitational. There is
no contradiction with the usual Lorentz invariance of the quantum
vacuum, since the presence of a mass or any non uniform
gravitational field breaks the Lorentz symmetry.  This variation
of the  virtual pairs lifetime in a gravitational field is not an
{\em ad hoc} hypothesis, but an unavoidable consequence of the
fourth Heisenberg relation. Note also that the virtual particles
considered here are not created by gravity: they are just the
usual zero point  particles that fill the space everywhere,
according to elementary quantum physics: they live a bit longer
(or shorter), that is all.

In the following we will deal with the variation of the observed values of certain quantities
between a spacetime point of the universe with potential $\Phi$ and a terrestrial observatory.
As they will be expressed at first order in the potential,  $(\Phi -\Phi _\oplus)$ will be
written instead of $\Phi$, $\Phi _\oplus$ being the present gravitational potential of all the universe here at Earth.

{\bf Vacuum permittivity and permeability in a gravitational
potential}. Following the previous considerations, we will admit
as a phenomenological hypothesis that the quantum vacuum can be
considered as a substratum, similar to an ordinary transparent
optical medium and characterized by a permittivity and a
permeability that depend on $\Phi$. As $\Phi$ becomes more
negative, its density increases and the bare electron charge
$e_{\rm bare}$  is renormalized further or, otherwise stated, the
observed charge must become smaller. In such a view the
permittivity and the permeability of the quantum vacuum  must
change to $\epsilon _{\rm r} \epsilon _0$ and $\mu _{\rm r}\mu
_0$, the first factor expressing the effect of its thickening (or
lightening). As we assume a weak field, we can express the
relative permittivity and permeability at first order in the
potential. Since both are equal to 1 now at Earth, their value at
a spacetime point with potential $\Phi$ can be expressed as:
\begin{equation}
\epsilon _{\rm r}=1-\beta (\Phi -\Phi _\oplus )/c^2,\;\;\;\; \mu _{\rm r}=1-\gamma (\Phi -\Phi _\oplus )/c^2,
\label{30}
\end{equation}
 $\beta$ and $\gamma$ being certain coefficients, which must be positive since the quantum vacuum is dielectric but paramagnetic
(its effect on the magnetic field is due to the magnetic moments of the virtual pairs).  It must be stressed that eqs. (\ref{30}) are a plausible hypothesis, consequence of the fourth Heisenberg relation.

 Because of (\ref{30}) the observed values of
the  electron charge and light velocity at potential $\Phi$ must be equal to $e/\epsilon _{\rm r}=e(1+\beta (\Phi -\Phi _\oplus )/c^2)$ and $c/\sqrt{\epsilon _{\rm r}\mu _{\rm r}}=c[1+(\beta +\gamma )(\Phi -\Phi _\oplus )/2c^2]$, at first order (the more negative is $\Phi$, the smaller the charge and the slower the light). Consequently $\alpha=e^2/4\pi \hbar c\epsilon _0$ must change to $\alpha \sqrt{\mu _{\rm r}/\epsilon _{\rm r}^3}$. This means that the change in the observed value of $\alpha$, as computed at Earth from the light absorbed or emitted at a spacetime point with potential $\Phi$,  must  be
\begin{equation}
\Delta \alpha /\alpha =\xi (\Phi -\Phi _\oplus)/c^2,
\label{40}
\end{equation}
 where $\xi = (3\beta -\gamma)/2$ and $\Phi _\oplus$ is the potential at Earth.

{\bf On the gravitational potential of all the universe}. As said before, this work considers only  weak gravitational potentials that verify $|\Phi |/c^2\ll 1$ and can be studied therefore in the Newtonian limit \cite{Pee93}. When that condition is no longer verified, the gravitational energy of a body is comparable to its rest energy and a relativistic theory must be used instead. This point, however, needs a clarification, since there are two independent different aspects in a Newtonian approximation. One is that the force on a particle must be weak and equal to $-\nabla \Phi$, the potential $\Phi$ being determined up to an additive constant, other is the value of the potential energy of a particle.  When studying a bounded system, as the Earth-Moon pair, a galaxy or a cluster, the effect of the faraway bodies can be neglected in the study of its motion, taking only the potential $\Phi$ caused by the nearby masses. The reason is clearly that the distant bodies contribute to $\Phi$ with a term that is almost independent of the space coordinates, at the scale of the system studied, and has therefore a negligible effect on the forces on the system.   This is correct for all practical purposes concerning the motion.
However, the gravitation being a long range force, one must be careful to include all the faraway matter that could have an effect on the problem {\em if something depends on its gravitational potential energy.  For this reason, the potentials $\Phi$ and $\Phi _\oplus$ that appear in eqs. (\ref{10})-(\ref{40}) will be taken in this work as being due to all the universe}. Note that this potential must include the contribution of matter, either ordinary or dark, as well as the effect of the cosmological constant.

Indeed the gravitational potential energy of a body is defined as the energy needed to bring it from the infinity to its actual position, without changing its kinetic energy or any other nongravitational energy. It could be argued that a body can not be brought actually from the infinity, since this is farther away than the horizon of the visible universe. However, it is clear that less energy is needed to create a virtual pair if $\Phi$ is negative than if it is zero, the difference playing the same role as the gravitational energy when bringing a body to a spacetime point at potential $\Phi$.
Therefore, the variation of the lifetime of the virtual pairs is due in this model to the gravitational interaction of the virtual pairs of the quantum vacuum with all the matter and radiation in the universe, and also with all the quantum vacuum itself.

It is assumed here that the universe is flat ({\em i.e.} $k=0$) and  consists in ordinary
 plus dark matter (with zero pressure) and dark energy due to the cosmological constant
  $\Lambda$. The densities over the critical density at present time are noted as
  usual $\Omega _M, \Omega _\Lambda$ (with $\Omega _M+\Omega _\Lambda=1$).  Taking
  $p_\Lambda =-\rho _\Lambda c^2$ as the equation of state of the quantum vacuum, the source
  of its gravity in the Newtonian approximation is $\rho_\Lambda +3p_\Lambda /c^2=-2\rho_\Lambda$.
  Consequently, the space average potentials at present time due to the matter and to the
  quantum vacuum are $\Phi _{\rm av,M}=\Omega _M\Phi _0$ and $\Phi _{\rm av, \Lambda} =-2\Omega _\Lambda \Phi _0$,
  respectively, where $\Phi _0/c^2=-\int _0^{R_U}G\rho _{\rm cr}4\pi r{\rm d}r/c^2 \simeq -0.3$
   is the potential that would be created by a mass distribution with the critical density,
   $R_U=3,000$ Mpc being taken as the visible universe radius.
   Most of these potentials is due to the larger distances. This means that the average
   gravitational potential produced by the visible universe is now equal to
   $\Phi _{\rm av}=\Phi _0(\Omega _M-2\Omega _\Lambda )$. The effect of the inhomogeneities can
   be neglected. For instance, the contributions to the potential of the Sun, of Earth and of
   the Galaxy at a terrestrial laboratory are, respectively, $\simeq - 10^{-8}c^2$, $\simeq -7\times 10^{-10}c^2$
   and $\simeq -6\times 10^{-7}c^2$, which are much weaker than $\Phi _{\rm av}$.

{\bf Cosmological variation of $\alpha$}. Webb {\em et al} measured lines absorbed by distant gas clouds at
high redshift, so that $\Phi$ in (\ref{40}) must be the potential
at the absorption time $t$, which can be taken to be approximately uniform, except for small scale inhomogeneities.  In the past, when the universe was more compact and
dense, the distances varied as the scale factor $a(t)$, the mass density as $a(t)^{-1}$ while the density of the quantum vacuum was constant (except for its variation due to eq. (\ref{20}) that would give a second order effect). Consequently, the space average potential at time $t$ was $\Phi _{\rm av}(t)=\Phi _0(\Omega _M/a(t)-2\Omega _\Lambda a^2(t))$. Introducing this expression in eq. (\ref{40}) and neglecting the local inhomogeneities, it turns out that $(\alpha _z-\alpha )/\alpha$,  can be expressed as
\begin{equation}
{\Delta \alpha \over \alpha} ={\xi}
{\Phi _0\over c^2}\left[\Omega _M \left({1\over a(t)}-1\right)-2\Omega _\Lambda \left(a^2(t)-1\right)\right].
\label{50}
\end{equation}
Assuming a flat dust universe, the time evolution of the scale factor is
\begin{equation}
a(t)=\left({\Omega _M\over \Omega _\Lambda }\right)^{1/3}\sinh ^{2/3}\left[{(3\Lambda )^{1/2}t\over 2}\right],
\label{60}
\end{equation}
with $\Lambda =8\pi G\Omega _\Lambda \rho _{\rm cr}$ (eq.
(\ref{60}) reduces to $a(t)=(t/t_0)^{2/3}$ if $\Lambda =0$, $t_0$
being the age of the universe). Equations (\ref{50})-(\ref{60})
give the main result of this work: the time dependence of
$\Delta\alpha /\alpha$. Note that, since $\xi \Phi _0<0$,  $\Delta
\alpha <0$ for $t<0$, so that the quantum vacuum must have been
optically denser in the past (the electron charge was more
renormalized than now since $\Phi _{\rm av}(t)-\Phi _{\rm
av}(t_0)<0$).

 The thick line in Figure \ref{fig1} shows the relative change of the fine structure
constant given by eqs. (\ref{50})-(\ref{60}) in the case $(\Omega _M=0.3, \Omega _\Lambda =0.7)$ as compared with the
observations by Webb {\em et al} \cite{Web01} (in units of $10^{-5}$)
versus the look back-time (in units of the age of the universe),
with $\xi =1.3\times 10^{-5}$. This is the value that gives the best fit
 (remind that $\Phi _0/c^2\simeq -0.3$); it was  obtained by minimizing
$\chi ^2$, the minimum value being $0.63$ per point. The thin line
shows the same result for  $(\Omega _M=1, \Omega _\Lambda =0)$.
The best fit was obtained here for $\xi =1.9\times 10^{-5}$, the
value of $\chi ^2$  being of $0.56$ per point. Although the latter
gives a slightly better fit, both curves fit well the
observational points taking into account  the large error bars,
their difference being small. As explained before, the light
velocity must be affected also by the quantum vacuum and be time
dependent. A similar argument shows that the refractive index at
time $t$ is $n(t)=c/c(t)=1-(\beta +\gamma )\Phi _0[\Omega _m
(1/a-1)-2\Omega _\Lambda (a^2-1)]/2c^2\, (>1)$, so that the light
velocity is an increasing function of time.

{\bf Validity of the model.} Being based on a first order approximation in $\Phi$, this model is valid only if $\Omega _M|\Phi _0|/a(t)c^2\ll 1$, what means that the gravitational potential energy of a particle is much smaller than its rest energy. This condition is verified at present time, approximately at least, but looses progressively its validity when $a(t)$ decreases so that the model can not be extended arbitrarily towards the past (in its present nonrelativistic formulation).

It must be stressed that the singularity of (\ref{50}) at $t=0$ has no physical meaning since it is outside of the applicability of the model. Neither it can be said that $\alpha$ is predicted to change too rapidly or to vanish near time zero. One can not apply these ideas, therefore, to the recombination era, which happened at a very high redshift. But the interval of validity does cover the range of the observations by Webb {\em et al}, at least as a good approximation. To extend this model farther away in the past, it must be reformulated in relativistic terms.

This is not to say that the Newtonian approximation can not be
used farther away in the past, for higher redshift. It can
certainly be used whenever it is not necessary to consider the
gravitational interaction of a system with all the universe.

{\bf On the Oklo and other data.}  Damour and Dyson \cite{Dam96}
analyzed the data from the natural reactor which operated 1.8
billion years ago at Oklo (Gabon) and concluded that the relative
change of $\alpha$ from then to now is in the interval
$(-0.9\times 10^{-7}, 1.2 \times 10^{-7})$ (assuming that other
constants like the Fermi constant do not vary). Equation
(\ref{50}) gives for that time the relative change $\Delta\alpha
/\alpha\simeq -1.3\times 10^{-6}$ (resp. $\simeq -5.4\times 10^{-7}$) if $\Omega _M=0.3,\Omega _\Lambda =0.7$
 (resp. $\Omega _M=1, \Omega _\Lambda =0$), values which are outside but not far
from that interval.

The Oklo study is thought to give the most powerful method to
determine the variation of $\alpha$ with geochemical data but,
according to Uzan \cite{Uza02} ``one has to understand and to
model carefully the correlations of the variations of $\alpha
_{\rm w}$ and $g_{\rm s}$ as well as the effect of $\mu$
($=m_e/m_p$). This difficult but necessary task remains to be
done".
 There are several others studies which set bounds on the variation
 of $|\Delta \alpha|$, using a number of different data \cite{Uza02}.
The results of the present model  are compatible with these other
bounds, except for the one set by Olive {\em et al} using data of
$\beta$-decay in meteorites \cite{Oli02}, $|\Delta \alpha /\alpha
|<3\times 10^{-7}$ during the past 4.6 Gyr (redshift about 0.45),
while this model gives $\simeq 3 \times 10^{-6}$  (resp. $\simeq
1.7 \times 10^{-6}$) if $\Omega _M=0.3,\Omega _\Lambda =0.7$
(resp. $\Omega _M=1, \Omega _\Lambda =0$). However,  this bound by
Olive {\em et al} ``can also be altered if the neutrinos are
massive", according to Uzan \cite{Uza02}.

In any case, this model gives a fair account of the observations by Webb {\em et al}.
Taking everything into account, it seems worth of consideration to explore its consequences.

{\bf Comparison with the gravitational redshift}.  The effect
described here produces a frequency shift which may seem similar to the
gravitational redshift $\Delta \omega /\omega = \Delta \Phi /c^2$.
The two effects are different, however. The one observed by Webb
{\em et al} is due to the variation of the fine structure
constant. The principal part of the frequencies emitted by an atom
is proportional to $(e^2/4\pi \hbar \epsilon _0)^2$, not exactly
to $\alpha ^2$ while  the relative separation between lines in the
relativistic fine structure splitting does depend on $\alpha ^2
=(e^2/4\pi \hbar c\epsilon _0)^2$.   On the other hand, the
gravitational shift  affects equally to all the lines in a
multiplet. In experiments that disregard the width of the
multiplets, the contribution of the change of $\alpha$ to the
frequency shift is obtained from (\ref{40}), but with $2\beta$
instead of $\xi$, {\em i.e.} $\Delta \omega /\omega = 4\beta
\Delta \Phi /c^2$. In that case, the observed shift would be
$\Delta \omega /\omega =(1+4\beta)\Delta \Phi /c^2$, {\em i.e.}
the addition of the two effects.  The best confirmations of
the gravitational redshift, those by Pound, Rebka and Snider,
agree with the prediction of General Relativity up to about 1 $\%$
\cite{Wei73}, but they refer to nuclear levels in which the
electromagnetism plays only a part. This means that the results of this work are
certainly compatible with the experiments on the gravitational redshift if
$\beta \leq 2.5 \times 10^{-3}$, an inequality that this work
assumes to be satisfied. However, there is a problem:  as $\xi <
3\beta /2$, a bound on $\beta$ is also a bound on $\xi$ but the
converse is not true. In any case, a necessary condition for the
compatibility of the results of this work and the gravitational
redshift experiments is that $\xi <4\times 10^{-3}$, while this model predicts that $\xi$ is equal to about $ 1.3 \times 10^{-5}$ and $1.9\times 10^{-5}$ in the two cases considered.

{\bf Two final comments}.
 First, note that the dressed
electron charge and the fine structure constant are not treated
here as universal constants but as the result of the interaction
of point charges with the quantum vacuum, their variations being
an indirect consequence of the universal expansion, through the
modification of their renormalization.

Second, because of the conservation of the charge conjugation, the pairs are created with $L=S=0$, so that their energy in a magnetic field is  $E-2\mu _BB$ with
probability 1/2, $E$ being the non-magnetic energy. For magnetic field close to $B_0 = mc^2/\mu _B\simeq
8.8\times 10^{13}$ gauss, some pairs would have zero total energy, their lifetime being infinite according to
the fourth Heisenberg relation).
If $B$ approaches $B_0$ from below, the creation of the pairs would begin to be energetically free. There would
be a threshold to some peculiar phenomenon, although it is not clear what this can be (note that the inequality $B\ll B_0$ is
a necessary condition for the validity of the Euler-Heisenberg Lagrangian to study in QED the
interactions of the quantum vacuum with a magnetic field \cite{Zav96}).
 Intriguingly, it turns out that $B_0$ is close to the highest magnetic field measured in magnetars, see Kouveliotou {\em et al} \cite{Kou98}. Commenting ref. \cite{Kou98}, Kulkarni and Thompson \cite{Kul98} say, ``our failure to detect radio pulsars with magnetic fields greater than $2\times 10^{13}$ gauss is because, at such high fields, quantum electrodynamic effects help to damp the radio emission" this being ``direct evidence of magnetic fields strong enough to perturb the very structure of the vacuum". This gives support to the present work:  the upper limit for the magnetic field of magnetars could be just a manifestation  of the fourth Heisenberg relation similar to the one considered here.

{\bf Summary and conclusions}. In the phenomenological model presented here, the cosmological variation of the fine structure constant is due to the combined effect of the fourth Heisenberg relation and the gravitational interaction of the virtual pairs in the zero-point radiation with all the universe. The problem is studied in the Newtonian approximation.  The quantum vacuum is treated as an optical medium characterized by its relative permittivity and permeability that depend on the average gravitational potential of the universe. The model predicts that $\Delta \alpha /\alpha$  is proportional to $\{\Omega _M[a(t)^{-1}-1]-2\Omega _\Lambda [a(t)^2-1]\}$ ($a(t)$ being the scale factor). The results for the cases $\Omega _M=0.3,\Omega _\Lambda =0.7$ and $\Omega _M=1, \Omega _\Lambda =0$ agree with  the observations by
Webb {\em et al} \cite{Web01} on the
cosmological  variation of the fine structure constant,  as is seen in figure 1.  The argument goes as follows.

(i) Because of the fourth Heisenberg relation, the lifetime of the virtual pairs of the zero-point radiation depends on the  gravitational potential $\Phi$. This causes the permittivity and the permeability of the quantum vacuum, the observable  electron charge and the light velocity to depend also on $\Phi$. The consequent change of the observed fine structure constant is expressed at first order in terms of the average gravitational potential of the universe, and  of a parameter $\xi$  related to the renormalization effects of the quantum vacuum.

(ii) As the universe was
more compact in the past, its average gravitational
potential was more negative (or less positive). Consequently, the lines of the spectra
of distant quasars were absorbed with a more renormalized value
of the electron charge than now. As a result, $\Delta \alpha /\alpha$ is  given by eq. (\ref{50}), which is plotted in fig. 1. The agreement seems good. Note that, in this model, the optical density of the quantum vacuum increases towards the past and decreases along the universe history.

(iii) The model can not be extended arbitrarily towards the past, since it is based on a first order Newtonian approximation that is valid for the recent past, including the range of the observations by Webb {\em et al}, but is no longer applicable to higher redshift. In particular, it can never be applied to the recombination era. A relativistic approach should be followed in order to extend the model to the remote past.
Note finally that, according to this model, the light is also affected by the gravitational potential so that it was slower in the past.

  The conclusion of this paper is that the combined effect of the gravitation of all the universe and of the fourth Heisenberg relation on the density of the zero-point radiation and the corresponding cosmological variations of $\alpha$, $e$ and $c$ proposed here should be further investigated.

I am grateful to Profs. Claudio Aroca, Ana In\'es G. de Castro, Elo\'{\i}sa L\'opez, Juan M.
Us\'on and Jos\'e L. Trueba for discussions.

\newpage

\begin{figure}
\vspace{5cm}
\caption{$\Delta \alpha /\alpha$ (times $10^{-5}$) vs fractional
look-back time, predicted by this work in the cases $\Omega _M=0.3,\Omega _\Lambda =0.7$ (eqs. (\ref{50})-(\ref{60}), thick line) and $\Omega _M=1,\Omega _\Lambda =0$ (thin line),   as compared
with the data by Webb {\em et al} \cite{Web01} (explanation in the text).}
\label{fig1}
\end{figure}

\end{document}